\documentclass[twocolumn,aps,prl]{revtex4}
\usepackage{graphicx}
\usepackage{times}
\usepackage{amsmath}
\usepackage{amsfonts}
\usepackage{amssymb}
\usepackage{units}
\usepackage{stmaryrd} %% "// sign" = \sslash
\DeclareGraphicsExtensions{.pdf,.png,.eps,.jpg}

\begin{document}

% Title of the article
\title{Controlling edge states in the Kane-Mele model via edge chirality}

% Authors
\author{%
Gabriel Aut\`es and Oleg V. Yazyev
}

% author's affiliations/addresses
\affiliation{%
Institute of Theoretical Physics, Ecole Polytechnique
F\'ed\'erale de Lausanne (EPFL), CH-1015 Lausanne,
Switzerland}
\date{\today}

\begin{abstract}
% This is a macro for the typesetting of two-column text in an
% abstract. It will typeset the two arguments in \abstcol{}{} as the
% left and right column inside the abstract box. At the
% columnbreak there will be always a columnbreak (\par), so both
% columns start with a new paragraph. No automatic column height
% balancing is done.
%
% If used with a \titlefigure it will silently output both
% parameters as consecutive paragraphs.
%
% The macro is defined exclusively inside the argument of \abstract{};
% if used outside it will raise an error.
%
% Usage: \abstcol{<left column>}{<right column>}

We investigate the dependence of band dispersion of the quantum spin Hall effect (QSHE) edge states in the Kane-Mele model on crystallographic orientation of the edges. Band structures of the one-dimensional honeycomb lattice ribbons show the presence of the QSHE edge states at all orientations of the edges given sufficiently strong spin-orbit interactions. We find that the Fermi velocities of the QSHE edge-state bands increase monotonically when the edge orientation changes from zigzag (chirality angle $\theta = 0^\circ$) to armchair ($\theta = 30^\circ$). We propose a simple analytical model to explain the numerical results.
\end{abstract}

\maketitle

% The class file requires the standard graphicx Latex package. See the 'LaTeX
% standard graphics and color packages documentation' for more information at
% <http://tug.ctan.org/tex-archive/macros/latex/required/graphics/grfguide.pdf>.
%
% Accepted figure file formats depend on which LaTeX flavour is used.
% Classic LaTeX is always able to use Encapsulated Postscript (EPS);
% PDFLaTeX can't use this but accepts PDF, JPG, PNG, and GIF formats.
%
% See examples for implementing graphics in floating figure environments later in this file.
% If \titlefigure is given, it takes as its mandatory parameter the
% name (without extension) of some figure file.

%\titlefigure[height=3.1cm]{empty2w}
%\titlefigurecaption{%
%  This is the caption of the \emph{optional} abstract figure. If
%  there is no abstract figure here, the abstract text should be divided into both columns.}

\maketitle   % please do not remove

In their seminal paper \cite{kan05-1}, Kane and Mele proposed a simple two-dimensional model which realizes the quantum spin Hall effect (QSHE). The model essentially considers a tight-binding model on a honeycomb lattice akin to graphene with added spin-orbit interactions. Realistic graphene is characterized by only very weak intrinsic spin-orbit coupling of the order of 10$^{-5}$~eV \cite{min06,yao07,boe07,gmi09,kon10}. While no QSHE has been experimentally observed in graphene, the Kane-Mele construction has become a popular model of $Z_2$ topological insulators \cite{Hasan10,Qi11}. Significant attention is currently devoted towards understanding the relationship between the crystallographic orientation of edges and surfaces in topological insulators and the resulting 
properties of topologically non-trivial boundary states \cite{Zhang12,Silvestrov12}.  

In this work, we establish a dependence between the band dispersion of the QSHE edge states in the Kane-Mele model and the crystallographic orientation of the edges. In particular, we show that 
Fermi velocities of the topological edge states strongly depend on edge orientation.

\begin{figure}%
\includegraphics*[width=\linewidth]{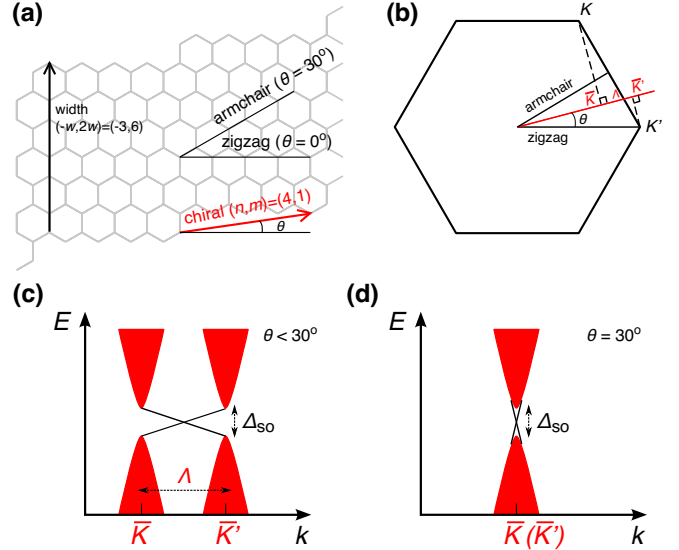}
\caption{%
(a) Atomic structure model of a chiral honeycomb ribbon characterized by the (4,1) edge translation vector ($\theta =10.9^\circ$) and width $w=3$. (b) Schematic illustration showing the projection of points $K$ and $K'$ of the 2D Brillouin zone of honeycomb lattice onto the direction corresponding to a chiral edge. Schematic band structures
of (c) zigzag and chiral ribbons ($0^\circ\le\theta< 30^\circ$), and (d) of armchair ribbons ($\theta=30^\circ$) in the presence of spin-orbit interactions. The quantum spin Hall effect edge states are shown as solid lines.}
\label{fig1}
\end{figure}

\begin{figure*}[t]%
\includegraphics*[width=\textwidth]{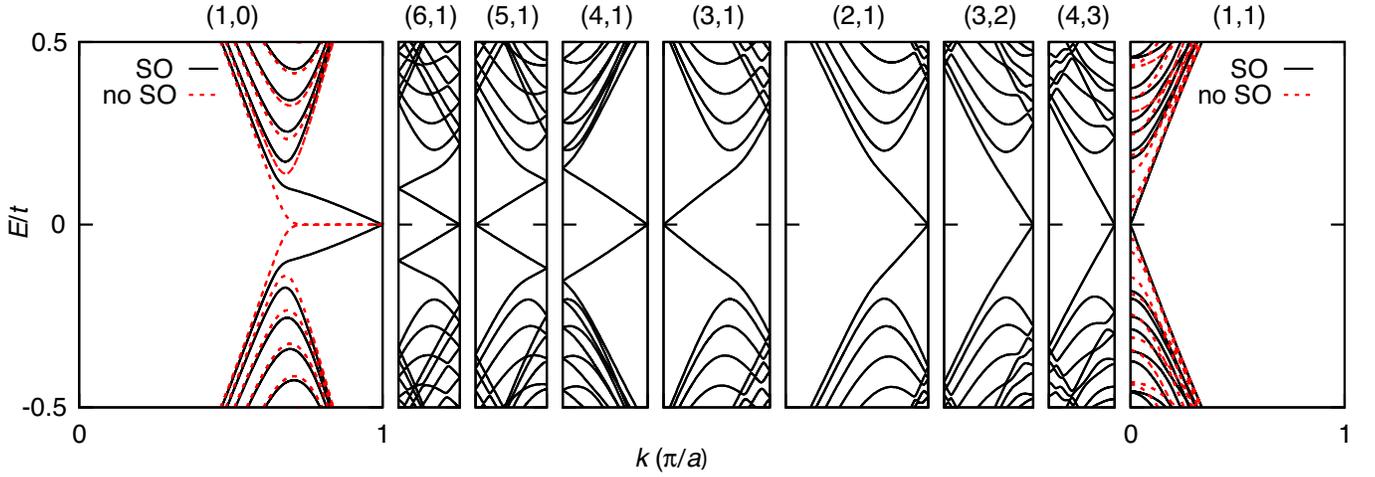}
\caption{Sequence of band structures of the honeycomb lattice ribbons
characterized by different periodicity vectors with widths defined by $w=16$ and $t_{\textrm{so}}=0.03t$. Band structures calculated in the absence of spin-orbit interactions are shown for the cases of zigzag ($\theta=0^\circ$) and armchair ($\theta=30^\circ$) nanoribbons. The scales of the plots account for the varying Brillouin-zone dimensions.
}
\label{fig2}
\end{figure*}

We investigate the electronic band structures of 1D periodic honeycomb lattice ribbons within the Kane-Mele model Hamiltonian \cite{kan05-1}
\begin{equation}
H=t\sum_{\left<i,j\right>,\sigma} c_{i\sigma}^{\dagger}c_{j\sigma} 
+ it_{\textrm{so}}\sum_{\left<\left<i,j\right>\right>, \sigma}  \nu_{ij} c_{i\sigma}^{\dagger} \sigma_z c_{j\sigma},
\label{eq1}
\end{equation}
where $\left<i,j\right>$ and  $\left<\left<i,j\right>\right>$ indicate first and second nearest neighbors, respectively, and $\sigma$ is the spin index. In this expression, the first term corresponds to an ordinary nearest-neighbor tight-binding model with hopping energy $t$. The second term introduces spin-orbit coupling of strength $t_{\textrm{so}}$. $\nu_{ij}=\pm 1$ is the Haldane factor \cite{hal88} defined as $\nu_{ij}=(\vec{d}_{ik} \times \vec{d}_{jk})/|\vec{d}_{ik} \times \vec{d}_{jk}|$ for a pair of second nearest neighbor sites $\left<\left<i,j\right>\right>$ connected via a common neighbor $k$. $\sigma_z$ is a Pauli matrix describing electron spin.
In graphene-like systems, the spin-orbit term opens a gap $\Delta_{\textrm{so}}=6\sqrt{3}t_{\textrm{so}}$ at the Dirac points. Following Kane and Mele \cite{kan05-1}, we choose the spin-orbit second neighbor hopping $t_{\textrm{so}}=0.03t$. We stress that the value used overestimates the intrinsic spin-orbit coupling present in realistic
graphene, which was predicted to be between 1 and 50~$\mu$eV according to first-principles calculations \cite{min06,yao07,boe07,gmi09,kon10}.

The configurations of investigated 1D periodic honeycomb lattice ribbons are defined by two parameters: (i) the crystallographic orientation of the edges and (ii) the width of the ribbon.
The edge direction is described by a translation vector $(n,m)$ of the graphene lattice (see Fig.~\ref{fig1}a). 
The high symmetry directions, armchair and zigzag, correspond to vectors $(1,1)$ and $(1,0)$, respectively. Equivalently, the 
edge orientation can be described in terms of chirality angle $\theta$
defined as the angle between the edge and the zigzag direction \cite{yaz11,tao11}. The edge translation vectors and chirality angles are 
related to each other by the following relation:
\begin{equation}
\theta=\arcsin\sqrt {\frac{3}{4}\left(\frac{m^2}{m^2+nm+n^2} \right)}.
\label{eq2}
\end{equation}
Following Ref.~\cite{yaz11}, we defined the width $w$ of the ribbon by the vector $(-w,2w)$ along the armchair direction as shown on Fig.\ref{fig1}a.

Figure~\ref{fig2} shows the band structures of honeycomb ribbons with chirality angles ranging from $\theta = 0^\circ$ ($(1,0)$ zigzag edge) to $\theta = 30^\circ$ ($(1,1)$ armchair edge) via a series of intermediate edge orientations (chiral edges). All considered models have comparable width defined by $w = 16$. In Figure~\ref{fig2}, one can immediately notice that all band structures feature linear band crossings occurring either at $k = 0$ or at the Brillouin-zone boundary $k = \pi/a$. The crossings display a clear increase of the Fermi velocity $v_{\rm F}$ upon increasing $\theta$. This relationship will be discussed in detail below. Analysis of
the electronic states at the band crossings reveals that the channels of opposite spins are localized at the opposite edges of 1D ribbon structures. That is, all investigated 1D honeycomb ribbons are in quantum spin Hall phase and exhibit spin-filtered edge states topologically protected against backscattering.

The effects of spin-orbit term are clearly illustrated for the case 
of a zigzag $(1,0)$ edge shown in Fig.~\ref{fig2}. In the absence of spin-orbit coupling ($t_{\textrm{so}}=0$; red dashed line) the band structure of the
ribbon model exhibits a dispersionless band at $E=0$. This band is four times degenerate (2 spins $\times$ 2 edges); it corresponds to edge-localized states originating from the lifted compensation between the two sublattices of the honeycomb lattice \cite{nak96}. The flat band connects $k= 2\pi/3a_0$ and $k= -2\pi/3a_0$ ($a_0$ is the lattice constant of the honeycomb lattice). These momenta correspond to the projections of points $K$ and $K'$ of the hexagonal Brillouin zone (the locations of the Dirac cones in the band structure of graphene) onto the momentum space of the 1D ribbon structures (points $\bar{K}$ and $\bar{K'}$ in Fig.~\ref{fig1}b). The introduction of spin-orbit 
term opens a band gap $\Delta_{\textrm{so}}$ at $\bar{K}$ and $\bar{K'}$ lifting the degeneracy of edge states and leading to a non-zero value of $v_{\rm F}$. In the quantum spin Hall phase the edge states connect the valence band at $\bar{K}$ with the conduction band $\bar{K'}$, and vice versa (Fig.~\ref{fig1}c).

The increase of edge chirality angle $\theta$ has a distinct effect on the electronic structure of honeycomb ribbons as it reduces the 
separation between points $\bar{K}$ and $\bar{K'}$
(see Figs.~\ref{fig1}b,c). More precisely, the distance between points $\bar{K}$ and $\bar{K'}$ is given by \cite{yaz11,Akhmerov08}
\begin{equation}
\Lambda=\frac{4\pi}{3a_0}\sin(\pi/6-\theta).
\end{equation}
This allows us to provide an estimate of the Fermi velocity as a function of spin-orbit interaction strength $t_{\textrm{so}}$ and chirality angle $\theta$:
\begin{equation}
v_F=\frac{\Delta_{\textrm{so}}}{\Lambda}=\frac{9\sqrt{3}t_{\textrm{so}}a_0}{2\pi\sin(\pi/6-\theta)}.
\label{vf}
\end{equation}

\begin{figure}[t]%
\includegraphics*[width=\linewidth,]{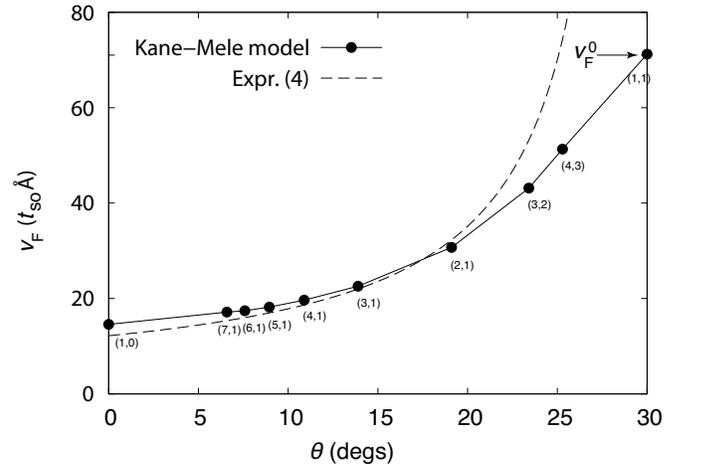}
\caption{%
Fermi velocity $v_{\rm F}$ of the quantum spin Hall effect edge states as a function of chirality angle $\theta$. The labels indicate the corresponding periodicity vectors of the edges.}
\label{fig3}
\end{figure}

Figure~\ref{fig3} compares the magnitudes of $v_{\rm F}$ obtained from band structure calculations performed on $w=20$ honeycomb ribbons, with the estimates provided by analytic expression (\ref{vf}). For $0^\circ < \theta < 20^\circ$, the analytic formula (\ref{vf}) shows very good agreement with the numerical results. 
As the chirality angle approaches the armchair edge limit, the 
computed values of $v_{\rm F}$ deviate from analytic estimates eventually resulting in a finite Fermi velocity at $\theta=30^{\circ}$.
The case of armchair edges is special as both $K$ and $K'$ are projected onto $k=0$ in the 1D ribbon band structure. The edge 
state dispersion in this situation is illustrated in Fig.\ref{fig1}d as well as in the calculated band structure in Fig.~\ref{fig2}. 
It follows that the Fermi velocity of the linear edge-state bands recovers the Fermi velocity of the massless Dirac fermions in the bulk when spin-orbit interactions are absent (the case of graphene), $v_{\rm F}^0=\sqrt{3}ta_0/2$ (Fig.~\ref{fig3}), confirming the recent
result of Gos\'albez-Mart\'inez {\it et al.} \cite{gos12}. Interestingly, this result does not depend on the strength of spin-orbit interactions, contrary
to the low-$\theta$ regime. On the other hand, armchair ribbons as well as high-$\theta$ chiral ribbons of finite width are semiconducting in the absence of spin-orbit interactions \cite{yaz11,son06}. Thus, spin-orbit coupling above certain critical  strength is required in order to bring these systems into the quantum spin Hall regime.

In summary, we investigated the dependence of the band dispersion 
of the topologically non-trivial edge states in the Kane-Mele model
on the crystallographic orientation of the edges. It was shown that the Fermi velocity of the quantum spin Hall edge states increases monotonically upon varying the edge chirality angle from $\theta =0^\circ$ (zigzag edge) to $\theta =30^\circ$ (armchair edge). 
A simple analytical model estimates the minimum 
Fermi velocity as $v_{\rm F} = 9\sqrt{3}t_{\textrm{so}}a_0/\pi$. The maximum value achieved for armchair edges recovers the Fermi velocity $v_{F}^0=\sqrt{3}ta_0/2$ of the Dirac fermions on honeycomb lattice 
in the absence of spin-orbit interactions. The relations established 
for this prototypical topological insulator provide an important insight into tailoring the properties of topologically protected boundary states in realistic materials.

We acknowledge support by the Swiss National Science Foundation (grant No. PP00P2\_133552).

\end{document}